\documentclass[%
 reprint,
superscriptaddress,
 amsmath,amssymb,
 aps,
]{revtex4-2}

\usepackage{graphicx}
\usepackage{dcolumn}
\usepackage{bm}
\usepackage{hyperref}

\usepackage{xcolor}
\usepackage{units}
\usepackage{amsmath}
\usepackage{amssymb}
\usepackage{mathtools} 
\usepackage{braket} 
\usepackage{graphicx}
\usepackage{soul}
\usepackage{tablefootnote}
\usepackage[makeroom]{cancel}
\usepackage[normalem]{ulem}
\usepackage{comment}
\usepackage{array}
\usepackage{ragged2e}
\usepackage{calc}

\newcommand{\oa}{\omega_\mathrm{a}}
\newcommand{\na}{n_\mathrm{a}}
\newcommand{\Na}{N_\mathrm{a}}
\newcommand{\Rp}{R_\mathrm{pump}}
\newcommand{\Pc}{P_\mathrm{c}}
\newcommand{\Ps}{P_\mathrm{s}}

\begin{document}

\preprint{APS/123-QED}

\title{Coherent polarization self-rotation}

\author{Roy Shaham}
\altaffiliation{Authors contributed equally.}
\affiliation{Department of Physics, Harvard University, Cambridge, Massachusetts, USA}
\affiliation{Department of Chemistry and Chem.~Biology, Harvard University, Cambridge, Massachusetts, USA}
\affiliation{Harvard-MIT Center for Ultracold Atoms, Cambridge, Massachusetts, USA}

\author{Orr Meron}
\altaffiliation{Authors contributed equally.}
\affiliation{Department of Physics of Complex Systems, Weizmann Institute of Science, Rehovot, Israel}
\affiliation{Rafael Ltd, Haifa, Israel}

\author{Or Katz}
\altaffiliation{Authors contributed equally.}\affiliation{School of Applied and Engineering Physics, Cornell University, Ithaca, New York, USA.}

\author{Dimitry Yankelev}
\affiliation{Rafael Ltd, Haifa, Israel}

\author{Ofer Firstenberg}
\email{ofer.firstenberg@weizmann.ac.il}
\affiliation{Department of Physics of Complex Systems, Weizmann Institute of Science, Rehovot, Israel}


\begin{abstract}
We introduce and study coherent polarization self-rotation (CPSR), a two-photon light-matter interaction in dense alkali-metal vapors that enables both narrowband optical spectroscopy of magnetic transitions and coherent coupling between light and collective atomic spins.
Unlike conventional polarization self-rotation, CPSR requires initial spin polarization and a predominantly linearly polarized probe.
It operates efficiently even in optically thick vapors with high buffer-gas pressure, rapid spin-exchange collisions, and optically-unresolved hyperfine structure.
We demonstrate CPSR with near-unity contrast in rubidium and achieve an exceptionally narrow two-photon linewidth of 10~Hz in potassium.
CPSR realizes a coherent interface between one optical quadrature and the long-lived collective electronic spin, offering a robust and scalable spin-light coupling in optically thick platforms.
This opens new opportunities for quantum optics, including quantum-enhanced metrology in the audio-frequency band and coherent transduction between light and ultra-long-lived noble-gas spins via alkali spins.
\end{abstract}

\maketitle

\section{\label{sec:intro}Introduction}

Optical spectroscopy and control of spin ensembles play central roles in both applied and fundamental science, from precision measurements and magnetometry to quantum information processing, where they enable tasks such as generating entanglement and quantum memories~\cite{Budker2013OpticalMagnetometryII, Sheng2013RomalisSubFemtoTesla, KorverWalker2015SynchronousSEOP, jiang2021BudkerALP,bloch2021nasduck,Julsgaard2001PolzikEntanglement, AlkaliNobleEntanglementKatz2020PRL,ThomasPolzik2021AlkaliMechanics}.
The performance of such spin-light interfaces hinges on two key properties: the coherence time of the collective spin and its optical coupling strength, which scales with the ensemble's optical depth.

Warm alkali-metal vapors provide a powerful platform for achieving both.
They exhibit strong optical transitions at accessible wavelengths and reach high atomic densities with moderate heating (50--200$\,^\circ$C).
Their technological maturity and modern, compact vapor-cell designs, make alkali vapors ideal for high-performance and miniaturized spin-based devices~\cite{Kitching2018atomicDevices,Lucivero2022LimesEarthGradiometer}.

Crucially, at elevated densities, warm alkali-metal vapors can enter the spin-exchange-relaxation-free (SERF) regime~\cite{Happer1977SERF,Savukov2005RomalisSERF}.
In this regime, the spin-exchange collision rate exceeds the Larmor frequency, locking the hyperfine and electron spin orientations to each other.
As the total spin orientation is conserved during spin exchange collisions, the coherence time of the hyperfine spin orientation is extended.
Its residual relaxation is caused by mixing of higher hyperfine moments, which still relax, due to non-vanishing Larmor precession.
The SERF regime is realized at high temperatures and low magnetic fields, and allows for long, tens of milliseconds, coherence times at high optical depths.

In practice, operating alkali vapor cells at elevated temperatures requires the use of buffer gases to suppress relaxation due to wall collisions.
However, this homogeneously broadens the optical transitions, making the hyperfine structure optically unresolved~\cite{Siddons2008RbSpectroscopy,Happer2010book,Romalis1997PressureBroadening,Katz2018storage1sec}.
This broadening undermines most multi-photon spectroscopic techniques, such as electromagnetically induced transparency~\cite{Fleischhauer2000LukinEITPolaritons} and nonlinear magneto-optical rotation~\cite{Budker2000NMOR}.
Furthermore, when alkali spins are used to mediate coupling to the nuclear spins of noble gases through their spin orientations, new spin-light interfacing schemes are required that couple to alkali spin orientation~\cite{Katz2022weak,Katz2022XstoragePRA,Shaham2021StrongCoupling,Katz2021NobleSpectroscopy}.

We introduce and experimentally demonstrate a two-photon spectroscopic method---coherent polarization self-rotation (CPSR)---that overcomes these limitations.
CPSR addresses the magnetic (Larmor) coherence of the collective spin in alkali vapors.
It operates effectively in dense, optically thick media with high buffer gas pressure, broadened optical lines, and fully unresolved hyperfine structure.
It addresses the spin orientation and therefore remains functional deep in the SERF regime and is immune to spin-exchange relaxation.
The mechanism combines optical Faraday rotation and vector lightshift.
Unlike standard polarization self-rotation (PSR)~\cite{Rochester2001PSR,Ries2003SqueezingPSR,bao2016spinSqueezing}, CPSR relies on externally prepared spin polarization and a predominantly linearly polarized probe. 
We demonstrate CPSR spectroscopic lines with near-unity contrast in rubidium, indicating efficient light-to-spin mapping, and 10 Hz-wide lines in potassium vapor.
CPSR establishes a platform for coherent light-matter interfaces that remain effective in high-density, high-temperature ensembles, enabling strong coupling to long-lived collective spin states and opening new directions in quantum sensing, optics, and hybrid systems involving noble-gas spins.

The paper is organized as follows:
Section \ref{sec:effect} introduces the CPSR mechanism.
Section \ref{sec:Rb} presents the rubidium experimental system, demonstrating CPSR absorption, amplification, and resilience to spin-exchange relaxation.
Section \ref{sec:10Hz} reports on CPSR with 10 Hz-wide lines in potassium.
Section \ref{sec:discussion} discusses performance limits, compares CPSR to PSR, and explores quantum applications.
The appendices include supporting data and modeling. 

\section{\label{sec:effect} Spin-light coupling mechanism}

\begin{figure}[t]
    \centering
    \includegraphics[width=1\columnwidth]{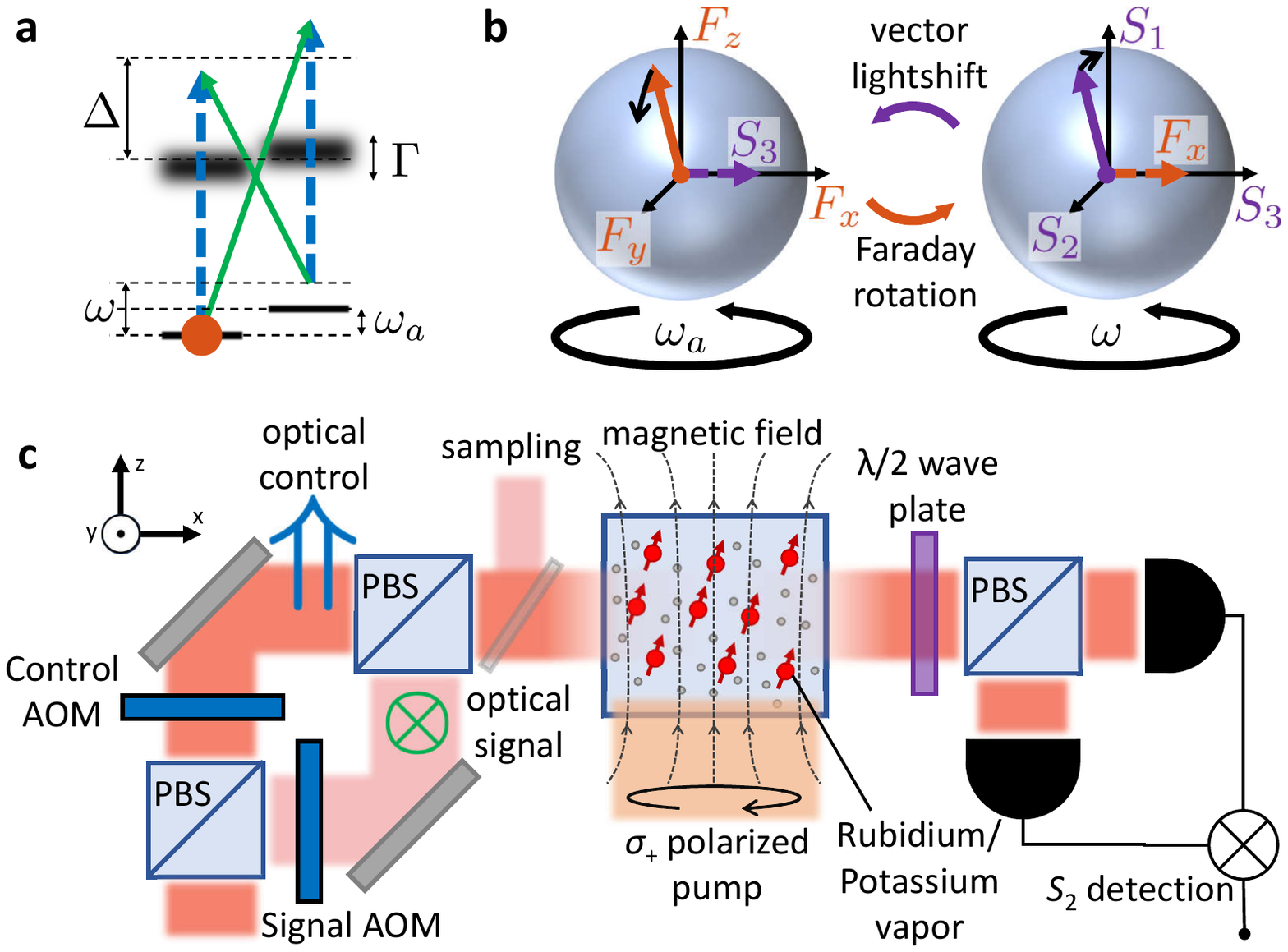}
    \caption{\textbf{Mechanism and setup for coherent polarization self-rotation.}
    \textbf{a}. In the high-temperature limit, where spin-exchange collisions are rapid, the alkali ensemble is reduced to an effective spin-1/2 system.
    The polarization-modulated optical probe decomposes into two photonic modes: a strong control (dashed, blue) and a weak signal (solid, green), which drive spin dynamics under the dispersive Faraday Hamiltonian.
    At the two-photon resonance ($\omega\approx\pm\oa$), the spin-polarized ensemble (red circle) undergoes coherent transitions within the ground-state manifold.
    \textbf{b}. Coupled precession of (left) the atomic spin on the Bloch sphere and (right) the optical polarization on the Poincaré sphere.
    Both are predominantly oriented along the vertical axis.
    Circular polarization ($S_3$) of the probe induces a vector lightshift that tilts the atomic spin, while the transverse spin component ($F_x$) rotates the optical polarization via the Faraday interaction.
    For $\omega\approx\oa$, this mutual coupling leads to destructive interference in the diagonal component ($S_2$), manifesting coherent two-photon absorption.
    \textbf{c}. Experimental setup.
    A warm vapor cell contains a dense alkali-metal spin ensemble (rubidium or potassium) and buffer gas.
    An optical-pumping beam (tan) polarizes the atoms, while a polarization-modulated probe, comprising strong control and weak signal modes, is sent through the cell.
    The coherent signal beam is generated by splitting light from the control beam using a polarizing beam splitter (PBS).
    Acousto-optical modulators (AOMs) adjust the amplitude, frequency, and relative phase of the two modes, introducing a kHz-scale frequency offset $\omega$.
    The outgoing $S_2$ component is measured using balanced polarization detection.
    }
    \label{fig:apparatus}
\end{figure}

We consider an alkali-vapor spin ensemble characterized by the collective hyperfine spin operators $\hat{\mathbf{F}}$.
These are defined as $\hat{F}_{\mu}=\sum_{j=1}^{\Na} \hat{F}^{j}_{\mu}$, where $\hat{F}^{j}_{\mu=\{x,y,z\}}$ are the hyperfine spin operators of the $j^\text{th}$ atom, and $\Na$ is the number of atoms within the probe beam.
Frequent spin-exchange collisions lead the spin populations to follow a spin-temperature distribution, such that $\hat{\mathbf{F}}$ effectively follows the electron spin, scaled by a factor $q$ known as the slowing-down factor~\cite{Allred2002RomalisSERFmagnetometer,Romalis1997Dissertation,AppeltHapper1998SEOPtheoryPRA}.
This reduces the multilevel ground-state manifold to an effective spin-1/2 system (see Fig.~\ref{fig:apparatus}a).
We further assume that the spins are optically polarized along the $z$ axis (Fig.~\ref{fig:apparatus}b, left), such that $\langle \hat{F}_z \rangle \gg \langle \hat{F}_x \rangle,\langle \hat{F}_y \rangle$, and $\langle\hat{F}_z\rangle=q \Na p_\mathrm{a}/2$, where $-1\le p_\mathrm{a}\le 1$ is the mean spin polarization~\cite{Walker1997SEOPReview,Happer2010book}.
An external magnetic field applied along $z$ induces Larmor precession at a frequency $\oa$, at the kHz range or below.

We consider an optical probe field detuned from the atomic transition by $\Delta\gg\Gamma$, where $2\Gamma$ is the full width at half maximum (FWHM) of the transition, including homogeneous (pressure) broadening due to collisions with buffer gas.
The probe propagates along the $x$ axis and is primarily polarized along $z$, with a small $y$ polarization component: $|\mathcal{E}_y|\ll|\mathcal{E}_z|$.
A frequency offset $\omega$ between $\mathcal{E}_z$ and $\mathcal{E}_y$, on the order of $\oa$, modulates the optical polarization between tilted linear and elliptical polarization.
This field can be viewed as a superposition of two linearly-polarized co-propagating photonic modes: the $z$-polarized mode acts as a strong control field (Fig.~\ref{fig:apparatus}, blue), while the  $y$-polarized mode serves as a weaker signal field (green).
These modes are detuned by the frequency $\omega$ and couple to different atomic transitions. 

We describe the light polarization state using the Stokes parameters: $S_1 \propto \frac{1}{2}|\mathcal{E}_z|^2-\frac{1}{2}|\mathcal{E}_y|^2\approx \frac{1}{2}|\mathcal{E}_z|^2$, $S_2 \propto \mathrm{Re}(\mathcal{E}_z^\ast \mathcal{E}_y$), and $S_3 \propto \mathrm{Im}(\mathcal{E}_z^\ast \mathcal{E}_y)$.
These are normalized such that $S_1$, $S_2$, and $S_3$ represent photon fluxes in the vertical, diagonal, and circular polarizations, respectively.
The time-dependent polarization traces a trajectory on the Poincaré sphere, precessing around the $S_1$ axis at frequency $\omega$ (see Fig.~\ref{fig:apparatus}b, right).
To leading order in $|\mathcal{E}_y/\mathcal{E}_z|$, $S_1$ characterizes the control field, while $S_{2,3}$ represent the signal quadratures relative to the control.

CPSR relies on a Faraday-type interaction between the signal field and the electron spins.
The light is far detuned from the optical atomic transition, such that the dispersive interaction dominates while absorption is suppressed~\cite{Braginsky1996QNDreview,Polzik2010ReviewRMP}.
The signal's $S_3$ polarization component induces a \emph{vector lightshift}, tilting the axial spin component $\hat{F}_z$ and generating a transverse component $\hat{F}_y$, which then precesses around the $z$ axis.
In turn, the transverse spin $\hat{F}_x$ arising from that precession acts back on the light via \emph{Faraday rotation}, tilting the optical polarization $S_1$ and interfering with the signal's $S_2$ component.
To first order in the interaction strength, the axial observables $\langle \hat{F}_z \rangle$ and $S_1$ are conserved.

We present in Appendix \ref{sec:model} both a detailed theoretical model for CPSR and a simplified analytic solution, derived under specific assumptions discussed therein. 
The simplified model captures the essential features of the spin-light dynamics and yields a set of coupled equations for the transverse spin, 
\begin{subequations}
\label{eq:in-out-model-1}
\begin{align}
    \partial_t \langle \hat{F}_x \rangle & = -\gamma \langle \hat{F}_x\rangle + \oa \langle \hat{F}_y \rangle, \\
    \partial_t \langle \hat{F}_y \rangle & = - \oa \langle \hat{F}_x \rangle -\gamma \langle \hat{F}_y \rangle - k_{{\mathrm{L}\rightarrow\mathrm{a}}}S_3^\text{in},
\end{align}
\end{subequations}
and for the signal field,
\begin{subequations}
\label{eq:in-out-model-2}
\begin{align}
    S_2^\text{out} & = e^{-d(\Delta)} ( S_2^\text{in} +  k_{{\mathrm{a}\rightarrow\mathrm{L}}} \langle \hat{F}_x \rangle ), \\
    S_3^\text{out} & = e^{-d(\Delta)} S_3^\text{in}.
\end{align}
\end{subequations}
Here, $S_{\{2,3\}}^\text{in/out}$ denote the polarization components of the signal before/after the cell, $\gamma$ is the total spin decoherence rate, and $d(\Delta)=\na \sigma(\Delta) \ell$ is the off-resonance optical depth, with $\na$ the atomic number density, $\ell$ the cell length along the $x$-axis, and $\sigma(\Delta)=\sigma(0)/(1+\Delta^2/\Gamma^2)$ the absorption cross section at detuning $\Delta$.
The coupling constants are given by 
\begin{equation}  
k_{{\mathrm{L}\rightarrow\mathrm{a}}}= \frac{1-e^{-d(\Delta)}}{d(\Delta)} 
\frac{2\Delta\sigma(\Delta)}{q\Gamma A}
\langle \hat{F}_z \rangle,~~~~k_{{\mathrm{a}\rightarrow\mathrm{L}}}= 
\frac{2\Delta\sigma(\Delta)}{q\Gamma A} S_1^\text{in}, 
\end{equation}
where $A$ is the beam cross-sectional area.
Light attenuation due to off-resonant scattering is included in the model, along with its contribution to the spin decoherence rate $\gamma$, which also accounts for other relaxation mechanisms.
The simplified model further assumes that the probe beam uniformly illuminates the entire cell volume, such that $\Na = \na A\ell$.

Two points are worth emphasizing.
First, $\sigma(\Delta)$ represents the detuned absorption cross section, with $\Gamma$ incorporating the pressure broadening from the buffer gas.
Second, the spin decoherence rate $\gamma$ includes a contribution $\gamma_\mathrm{se}$ from spin-exchange collisions.
When the spin-exchange collision rate $R_\mathrm{se}$ exceeds the Larmor frequency $\oa$, the system enters the SERF regime, in which $\gamma_\mathrm{se}$ drops below the spin-exchange collision rate $R_\mathrm{se}$, scaling quadratically with $\oa$, while higher hyperfine spin moments (\textit{e.g.}, spin alignment) relax~\cite{Happer1977SERF,Allred2002RomalisSERFmagnetometer}. 
Both our theoretical treatment and all experimental measurements in this work are carried out in the SERF regime.
Full expressions for $\gamma$ and $\gamma_\mathrm{se}$ in this limit are provided in Table~\ref{tab:parameters} in Appendix~\ref{sec:model}. 

We now explicitly introduce the frequency offset $\omega$ of the signal field, $S_2^\text{in}=S_\perp e^{i\omega t}$ and $S_3^\text{in}=iS_\perp e^{i\omega t}$, into Eqs.~(\ref{eq:in-out-model-1}) and (\ref{eq:in-out-model-2}).
Assuming that $\langle \hat{F}_z\rangle$ remains constant, we solve the coupled equations in Fourier space to obtain the outgoing components $S_2^\text{out}$ and $S_3^\text{out}$, yielding the following spectrum:
\begin{equation} \label{eq:spectrum}
\frac{S_2^\text{out}(\omega)}{e^{-d(\Delta)} S_2^\text{in}(\omega)}=1 - \frac{i\Omega \oa}{\oa^2-(\omega-i\gamma)^2}\equiv 1 - \mathcal{C}(\omega).
\end{equation}
Here, we define the complex contrast $\mathcal{C}(\omega)$ as the ratio between the Stokes parameter induced by the atomic medium and the transmitted input, and identify the spin-probe coupling rate $\Omega=k_{{\mathrm{L}\rightarrow\mathrm{a}}}k_{{\mathrm{a}\rightarrow\mathrm{L}}}$ (see Appendix~\ref{sec:model}).
Notably, $\Omega$ scales with both the number of $z$-polarized atoms $\langle \hat{F}_z \rangle$ and the input photon flux $S_1^\text{in}$.

In the limit of negligible probe absorption, \textit{i.e.} $|\Delta|\gg \Gamma$ and $d(\Delta)\ll 1$,  we find:
\begin{equation}
    \Omega\approx d(0) p_\mathrm{a} R_\mathrm{c}/q,\label{Eq:approxkappa}
\end{equation}
where $d(0)$ is the on-resonance optical depth, and $R_\mathrm{c}=2\sigma(\Delta)S_1^\text{in}/A$ is the spin relaxation rate due to absorption of control light.

We examine the spectrum in the limit $\gamma\ll\oa$ while still maintaining SERF.
On resonance $\omega=\sqrt{\oa^2+\gamma^2}\approx\oa$, we find that the $S_2$ component of the signal is absorbed, $|1 - \mathcal{C}|< 1$.
Full absorption occurs when $\Omega=(\omega/\oa)2\gamma\approx2\gamma$; 
for $\Omega> 2\gamma$, the $S_2$ undergoes a phase inversion.
At the anti-resonance $\omega\approx-\oa$, the $S_2$ component is amplified, $|1 - \mathcal{C}|> 1$.
These processes constitute two-photon transitions between effective Zeeman levels (see Fig.~\ref{fig:apparatus}a), with the absorption or amplification of the signal's $S_2$ component at $\omega\approx\pm\oa$ accompanied by coherent excitation of the collective transverse spin.

At the two-photon resonance, the atoms precess on the Bloch sphere with the modulation frequency of the Stokes parameters and follow their Poincaré trajectory with a $\pi/2$ phase delay (see Fig.~\ref{fig:apparatus}b).
Consequently, the atomic-induced $S_2$ response (the atomic readout) is $\pi$ out of phase with the input, leading to their destructive interference, measured as $S_2$ absorption.
At the anti-resonance, the precession frequency is opposite to the Stokes modulation, such that while $F_x,S_3$ behave similarly, $F_y$ mirrors its phase, leading to in-phase input and atomically-induced $S_2$, hence their constructive interference, measured as $S_2$ gain.
In fact, examination of $S_1$ up to second order in the interaction strength reveals that absorption (gain) in $S_2$ is accompanied by fine attenuation (amplification) of $S_1$, reflecting a coherent signal-to-control exchange.

\section{\label{sec:Rb}Rubidium CPSR spectroscopy}

The experimental apparatus consists of a magnetically shielded, 1-cm$^3$ cubic glass cell containing rubidium atoms at natural isotopic abundance (see Fig.~\ref{fig:apparatus}c).
The cell is heated to 154~$^\circ$C, yielding a rubidium density of $\na=1.22\times10^{14}~\mathrm{cm^{-3}}$, and a spin-exchange collision rate of $R_\mathrm{se}\approx 16$~kHz 
(here and throughout, $\mathrm{Hz}\equiv2\pi\cdot\mathrm{s}^{-1}$). 
The cell also contains 450 Torr of neon buffer gas, which renders atomic motion diffusive and suppresses alkali spin relaxation due to wall collisions.
An additional 40 Torr of nitrogen is used to non-radiatively quench the optical excited state, mitigating radiation trapping and enhancing optical pumping efficiency in this optically thick medium.
The D1 transition exhibits a homogeneous linewidth of $\Gamma=\unit[2.6]{GHz}$, dominated by collisional broadening from the buffer gases (see Appendix\,\ref{ss:onephotonline}).
We estimate the on-resonance optical depth to be $d(\Delta=0)=130$.

The atomic spins are subject to a magnetic field $B_z$ along the $z$-axis.
They are optically pumped and held polarized along the $z$ direction using an auxiliary, circularly polarized $\unit[795]{nm}$ laser beam.
The pumping beam, with power up to $\unit[100]{mW}$, covers the entire cell and is blue-detuned by 45~GHz from the D1 transition.
The large pump detuning is chosen to mitigate pump depletion while preserving high spin polarization.
Low pump depletion helps homogenize the spin polarization and decoherence, which is significant for the spectroscopic measurement.
We estimate the corresponding pumping rate to be $\Rp/q=23$~Hz.
The spin decoherence rate is measured in the absence of optical pumping at low Larmor frequency, $\gamma(\Rp=\oa=0)=21.5$~Hz.
This results in a spin polarization of $p_\mathrm{a}= 0.52$ and a slowing-down factor of $q=7.4$ (see Appendices\,\ref{appendix:characterization} and \ref{sec:model}).
 
The apparatus is irradiated with a polarization-modulated probe beam, formed by interfering two mode-matched polarization components---the control and signal---that co-propagate along the $x$-axis.
The probe light, at 795 nm, is blue-detuned from the D1 transition by $\Delta=90$ GHz.
Acousto-optical modulators (AOMs) apply a differential frequency shift $\omega$ between the two modes, producing the polarization-modulated field with tunable modulation frequency, phase, and amplitude.
The control mode has power $\Pc=\unit[15]{mW}$ and is polarized along the $z$-axis, aligned with the atomic spin polarization.
The signal mode is polarized along the $y$-axis, with power $\Ps=\unit[20]{\mu W}$.
Both beams illuminate the full cell cross-section, with area $A\approx (\unit[1]{cm})^2$.
Since $\Pc\gg \Ps$, the Stokes components satisfy $S_1^\text{in}\gg S_{2,3}^\text{in}$.
The outgoing diagonal polarization component $S_2^\text{out}$ is measured using balanced photodetection.

\begin{figure}[tb]
    \centering
    \includegraphics[width=1.0\columnwidth]{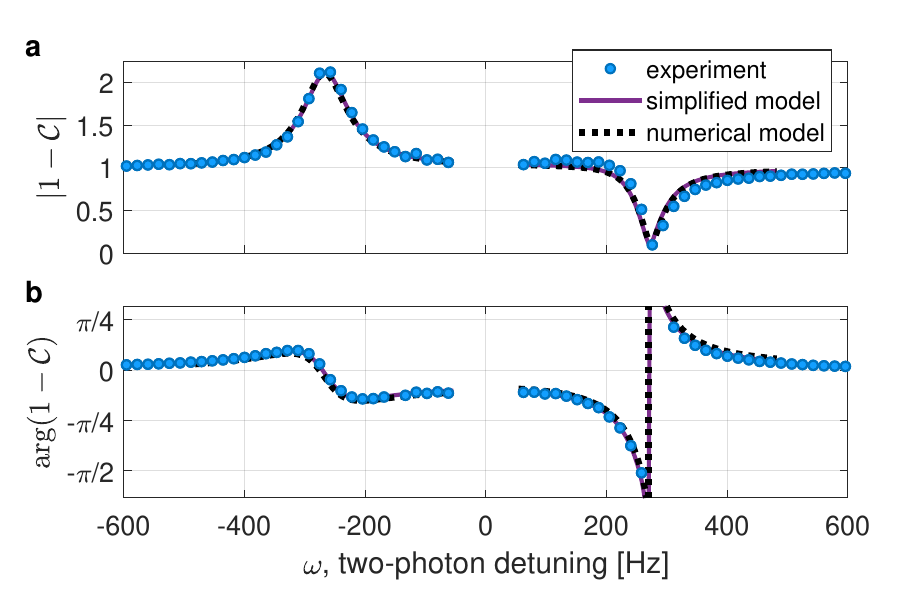}
    \caption{\textbf{A complex spectrum of coherent polarization self-rotation.}
    Measured transmission (\textbf{a}) amplitude and (\textbf{b}) phase of the optical signal traversing a rubidium vapor cell.
    The signal exhibits absorption at positive detuning and amplification at negative detuning, consistent with the predicted response near $\omega=\pm\oa$.
	These spectral features, with a contrast exceeding unity and a full width below 100~Hz, arise from the dispersive Faraday interaction between the polarized collective alkali-spin and the polarization-modulated probe.
    The numerical solution of the detailed theoretical model (dotted black line) and the simplified model from Eq.~(\ref{eq:spectrum}) (solid purple line) both agree well with the experimental data (circles).}
    \label{fig:spectrum}
\end{figure}

To demonstrate CPSR, we apply a magnetic field of $B_z\approx1$ mG, corresponding to a Larmor frequency of $\oa=268$~Hz including the pump-induced lightshift, and measure the transmission spectrum by scanning the polarization modulation frequency $\omega$.
The amplitude and phase of the outgoing quadrature $S_2^\text{out}$ referenced to those of the probe beam before the cell are plotted in Fig.~\ref{fig:spectrum}.
The spectra exhibit on-resonance absorption and anti-resonance amplification near $\omega=\pm\oa$, with a FWHM $\approx2\gamma$ of approximately 100 Hz, primarily limited by pumping-induced relaxation (see Appendix~\ref{appendix:characterization}).
The observed contrast reaches $|\mathcal{C}|=1.1$, exceeding unity, from which we infer a spin-light coupling rate of $\Omega=2\gamma|\mathcal{C}|\approx\unit[110]{Hz}$.
The near-complete extinction at $\omega=\oa$ highlights the potential for highly efficient, coherent mapping of the light field onto the collective atomic spin.

Numerical calculations based on the simplified model from Eq.~(\ref{eq:spectrum}) as well as the detailed theoretical model show excellent agreement with the experimental data, as demonstrated in Fig.~\ref{fig:spectrum}.
These calculations use independently measured physical parameters, with the exception of $\na$, $\gamma$, $\oa$, and $\Delta$, which are fine-tuned within the experimental uncertainty
(see Appendix~\ref{sec:model} for details on the model and parameter values).

\begin{figure}[tb]
    \includegraphics[width=1.0\columnwidth]{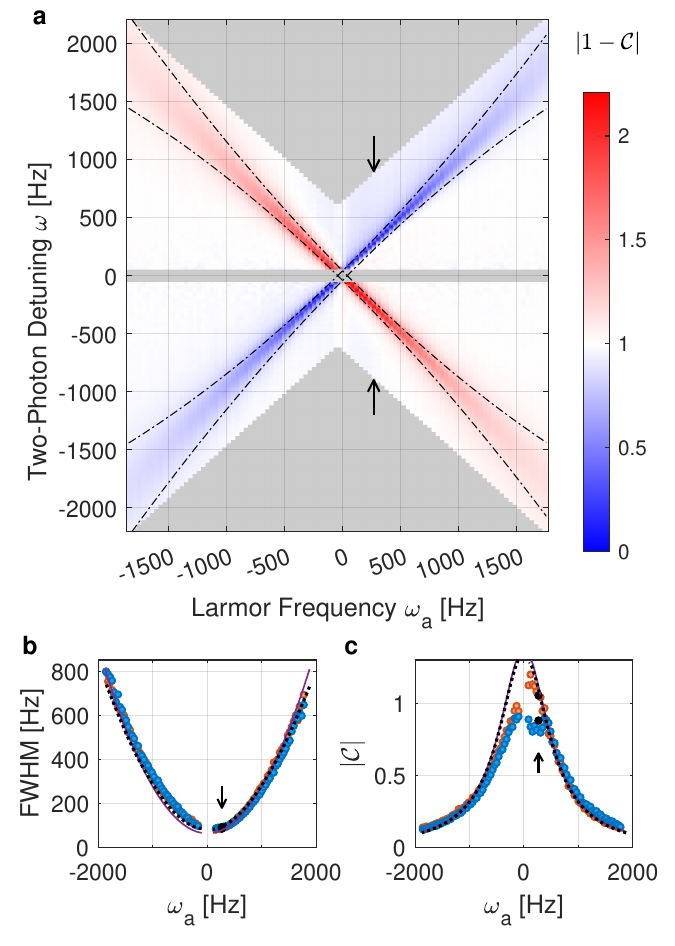}
    \caption{\textbf{Resistance of CPSR to spin-exchange relaxation.} 
    \textbf{a}. Transmission spectra of the optical signal for a range of Larmor frequencies $\oa$.
    Dashed lines indicate the fitted linewidths (FWHM) of the absorption (blue) and amplification (red) peaks.
    The spectral narrowing at low $\oa$ is a hallmark of the SERF regime, where Zeeman coherence is preserved despite frequent spin-exchange collisions. 
    Black arrows indicate the data shown in Fig.~\ref{fig:spectrum}.
    \textbf{b}. Dependence of the fitted FWHM on $\oa$, along with the theoretical prediction (black dotted line), showing the expected quadratic dependence characteristic of the SERF mechanism. 
    \textbf{c}. Contrast of the measured optical lines, 
    extracted from the transmission extrema.
    Contrast exceeding unity is observed for the amplification peaks at low $\omega$.
    In panels (b) and (c), dotted black and solid purple lines are the predictions of the detailed numerical model and simplified model [Eq.~(\ref{eq:spectrum})], respectively.}
    \label{fig:SERF}
\end{figure}

CPSR couples to the spin orientation and is therefore inherently resistant to spin-exchange relaxation in the SERF regime.
To demonstrate this, we measure transmission spectra across a range of Larmor frequencies satisfying $|\oa|\ll R_\mathrm{se}$ by varying $B_z$.
The results are shown in Fig.~\ref{fig:SERF}a. 
As expected, the CPSR spectral lines are significantly narrower than the spin-exchange collisions rate $R_\mathrm{se}$ and become progressively narrower as $|\oa|$ decreases.
Figures \ref{fig:SERF}b and \ref{fig:SERF}c summarize the extracted linewidth and contrast, respectively, as functions of $\oa$.
We observe a clear quadratic dependence of the linewidth on $\oa$, characteristic of SERF behavior, along with a reduction of line contrast as the relaxation rate increases.

Like other coherent spectroscopy techniques, CPSR requires auxiliary optical pumping for generating spin polarization, as reflected in the dependence of $\Omega$ on $p_\mathrm{a}$.
However, being a coherent process, it is also sensitive to spin relaxation induced by the pumping beam, particularly at high pump powers.
These competing effects create a trade-off between efficiency and coherence time.
Since spin polarization eventually saturates, this trade-off defines an optimal pumping rate for maximizing CPSR performance.
We analyze this effect and present the measured dependence of the CPSR contrast on the pumping rate in Appendix~\ref{ss:OP}.

These results demonstrate the feasibility of achieving high-contrast, narrow optical spectral lines in optically thick alkali vapor under high buffer-gas pressure, conditions traditionally considered challenging for coherent spin-light coupling.

\section{\label{sec:10Hz}Potassium CPSR spectroscopy with a 10-Hz linewidth}

The optical linewidth in CPSR is fundamentally limited by the coherence time of the electronic spin orientation and can therefore be made extremely narrow.
To demonstrate this, we performed CPSR measurements in a second setup using a 14 mm-diameter aluminosilicate glass sphere containing metallic potassium, 1500 Torr of helium, and 40 Torr of nitrogen.
Under these conditions, potassium exhibits a longer spin-orientation lifetime than rubidium, owing to its smaller spin-destruction cross sections~\cite{Happer2010book}.
However, the achievable CPSR contrast is reduced by the larger pressure broadening of $\unit[12]{GHz}$, caused by the high-pressure buffer gases.

At a temperature of 185~$^\circ$C, the potassium vapor density was $\na=\unit[7.7\cdot10^{13}]{cm^{-3}}$, yielding an on-resonance optical depth of $d(0) \approx 50$.
The atoms were optically pumped using circularly polarized light at $\unit[770]{nm}$ with a power of $\unit[13]{mW}$, blue-detuned by $\unit[100]{GHz}$ from the atomic transition.
A pump beam diameter of $\unit[4]{cm}$ ensured uniform pumping of the cell volume, reducing broadening of the CPSR lines.
The probe beam, red-detuned by $\unit[215]{GHz}$ from the same transition, had a power of $\unit[25]{mW}$.
For a Larmor frequency of $\oa=\unit[29]{Hz}$ (magnetic field of $B_z=60~\mu$G), the measured potassium spin coherence time was 50 ms.

\begin{figure}[tb]
    \centering     \includegraphics[width=0.9\columnwidth,clip]{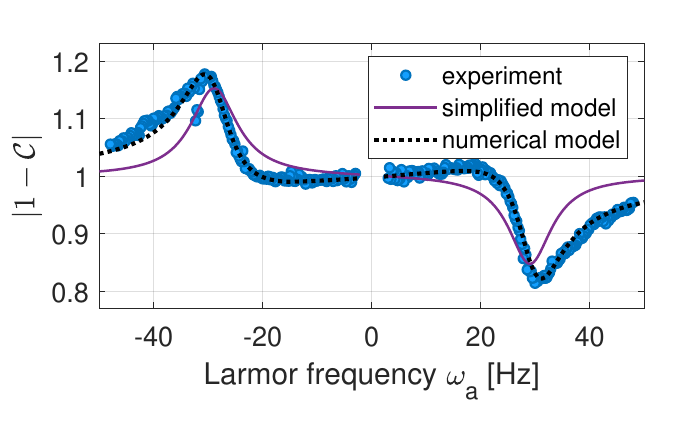}
	\caption{\textbf{CPSR in a spin-polarized potassium ensemble.}
    The transmission spectrum demonstrates CPSR resonances at $\pm\oa=\pm 29$~Hz with a full width of $2\gamma=9.7$~Hz and $\sim$20\% contrast.
    The data (blue circles) deviates from the simplified model's prediction [Eq.~(\ref{eq:spectrum}), solid purple lines] due to non-negligible optical pumping by the input $S_3$ component, which is captured by the numerical model (dotted black lines)} (see Appendix~\ref{sec:model}).
    \label{fig:10Hz}
\end{figure}

The corresponding CPSR spectrum, shown in Fig.~\ref{fig:10Hz}, exhibits a narrow optical linewidth of $\unit[9.7]{Hz}$, with a contrast of 20\%.
Notably, the resonance lineshapes deviate from the generalized Lorentzian form predicted by Eq.~(\ref{eq:spectrum}) due to optical pumping induced by the $S_3$ component of the probe itself.
The linewidth here is limited by this probe-induced pumping, as well as by depopulation following potassium-helium collisions.
While this effect is neglected in the simplified model (Eqs.~(\ref{eq:in-out-model-1})\,), it is fully captured by the detailed model presented in Appendix~\ref{sec:model}, which fits the data well.

\section{\label{sec:discussion}Discussion}

CPSR introduces a robust, versatile framework for quantum optics and spectroscopy in highly broadened, optically thick spin ensembles.
It operates effectively in the spin-exchange relaxation-free regime under high buffer-gas pressure and large optical depth, exhibiting narrow-band, near-unity coherent absorption or strong gain even under conditions that typically hinder other coherent schemes.
Performance is governed by the spin-light coupling rate $\Omega$ [given by Eq.~(\ref{Eq:approxkappa}) for weak probe absorption].
Maximal absorption at $\omega\approx\oa$, corresponding to unity contrast $\mathcal{C}=1$, occurs for $\Omega=2\gamma$, where $\gamma$ is the spin decoherence rate.
For larger $\Omega$, the coupling strength and amplification at $\omega\approx-\oa$ increase, with contrast reaching $|\mathcal{C}|=\Omega/(2\gamma)$.

The coupling can be increased by raising the control power $\Pc$ up to saturation, occurring since at high power both $\Omega\propto \Pc$ and $\gamma\propto \Pc$.
In our system, $|\mathcal{C}|$ would saturate near $\sim40$ for $\Pc>50$~mW at $30$~GHz pump detuning.
Further enhancement is possible by increasing the optical pumping rate and the cell temperature, provided that the pump power is sufficient to avoid depletion.
If depletion becomes significant, spatial effects such as inhomogeneous light shifts must also be accounted for.
Miniaturized vapor cells (\textit{e.g.}, 10 mm$^3$) offer a route to compact, high-performance CPSR applications.
In rubidium, under feasible conditions (\textit{e.g.}, 10~mW laser power, 250~$^\circ$C cell temperature, and 1000~Torr buffer gas pressure), CPSR could reach $|\mathcal{C}|>7$ with a linewidth on the order of $500$~Hz.

While CPSR benefits from the long coherence times and high optical depth of the SERF regime and exhibits magnetically sensitive spectral features, it differs conceptually from conventional SERF magnetometers.
In SERF magnetometry, spin orientation is commonly excited magnetically and read out optically via Faraday rotation.
In CPSR, both excitation and readout are optical, driven by vector lightshifts, forming a two-photon interface that generates the observed optical spectra.
Consequently, SERF magnetometers typically operate in the weakly coupled probe regime, where the spin coherence is minimally perturbed by the probe, whereas CPSR relies on strong, bidirectional spin-light coupling.
Optical driving also enables spatial resolution set by the beam size rather than global magnetic excitation, albeit with reduced sensitivity at small spatial scales due to atomic thermal motion.

CPSR relies on the dispersive light-matter interaction, which preserves coherence between the optical field and the collective spin.
This differs from PSR, in which spin orientation is generated by dissipative optical pumping via the circular component of the elliptically polarized beam, and the resulting polarization rotation arises from dispersive Faraday rotation.
The dissipative optical absorption in PSR limits the realization of a fully coherent light-matter interface.
Accordingly, CPSR requires an auxiliary pump to establish spin orientation.
Its coupling strength increases with the on-resonance optical depth $d(0)$ [$\Omega\propto d(0)$ in Eq.~(\ref{Eq:approxkappa})], whereas PSR exhibits an optimal optical depth near $d(0)\approx1$, with performance degrading as $d(0)$ increases.

Notably, since CPSR is both coherent and optically driven, it realizes the \emph{Faraday Interaction} (Quantum-Non-Demolition, QND) Hamiltonian~\cite{Braginsky1996QNDreview}.
Optical fields can carry nonclassical states, thereby enabling a quantum interface with the collective spin and opening avenues for quantum information applications.
In Appendix~\ref{subsec:operators}, we show that the input-output relations extend to quantum operators [Eq.~(\ref{eq:nonclassical-FPOR-photons})].
CPSR enables nonclassical protocols previously demonstrated in spin-preserving-coated cells~\cite{Polzik2010ReviewRMP} to be implemented in buffer-gas SERF cells.
The key distinction is the presence of rapid spin-exchange interactions and strong optical broadening.
In earlier demonstrations, atomic interactions were considered a source of decoherence and mostly neglected, focusing on the collective coupling to the optical field~\cite{Borregaard2016SinglePhotonsOnMotionallyAveragedMicrocellsNcomm}.
In contrast, SERF enables long coherence times and high optical depth without specialized coatings.

We emphasize that CPSR is not intended to outperform state-of-the-art SERF magnetometers in magnetic-field sensitivity.
These have demonstrated sensitivity at the spin projection noise limit and QND readout~\cite{Vasilakis2011RomalisBackactionEvation, Shah2010highBWmagnetometryAndAnalysis}, and their exceptional sensitivity has been achieved through careful optimization of experimental parameters, such as increasing optical depth~\cite{Sheng2013RomalisSubFemtoTesla}.
CPSR, on the other hand, enables the use of collective spin orientation as a quantum resource for quantum information and precision measurement applications.

As near-term milestones, CPSR in the nonclassical regime enables (i) generation of nonlocal entanglement between two spatially separated ensembles via QND measurement of a joint spin quadrature~\cite{Julsgaard2004SqueezingStorage}, and (ii) a double-pass geometry that implements a spin-light beam-splitter Hamiltonian for efficient storage and retrieval of weak optical fields~\cite{Muschik2006storage}, with extensions toward on-demand single-photon generation~\cite{Dideriksen2021PolzikSinglePhotonMemory}.

These milestones open a path to quantum-enhanced metrology in the acoustic band.
LIGO is a particularly significant test case: it is one of the few platforms where quantum enhancement has already delivered system-level sensitivity gains.
It operates in the acoustic band, where the SERF regime combines high optical depth with suppression of spin-exchange-induced decoherence, and back-action-evasive readout via QND (negative mass reference frame) has been proposed~\cite{MollerPolzik2017qBAEnegativemass}.
Recent advances, including fast, sub-cycle squeezing~\cite{BaerentsenPolzik2024FastSqueeze}, and wavelength transduction of squeezing~\cite{NovikovPolzik2025AcousticNetwork}, outline a viable path to acoustic band implementations~\cite{JiaPolzik2025AcousticDemo}, for which CPSR in the SERF regime is well suited.

A practical consideration for such protocols is spin polarization.
Moderate nonclassical performance typically requires $\gtrsim70\%$ polarization~\cite{Katz2022weak}, whereas $>95\%$ is readily achievable in alkali vapors~\cite{ChenWalkerBabcock2007Polarization}.
Pump-induced relaxation can be mitigated by shutting the pump off during the quantum sequence.

Finally, CPSR provides fast, on-resonance quantum-optical coupling to long-lived nuclear spins of noble gases, via a dense alkali vapor.
Whereas prior work addressed noble-gas nuclei via weak, adiabatic alkali excitation to resolve mHz resonances~\cite{Katz2021NobleSpectroscopy}, CPSR supports substantial coherent alkali excitation, unlocking the strong-coupling regime~\cite{Shaham2021StrongCoupling}.
Strong coupling could facilitate orders-of-magnitude improvement of light storage bandwidth~\cite{Katz2022XstoragePRA,Katz2024OptimalStorageArxiv}.

\begin{acknowledgments}
We thank Mark Dikopoltsev for help with the experimental setup.
We acknowledge financial support by the Minerva Foundation, the Leona M.~and Harry B.~Helmsley Charitable Trust, the Shimon and Golde Picker - Weizmann Annual Grant, and the Laboratory in Memory of Leon and Blacky Broder.
\end{acknowledgments}


\appendix

\section{Auxiliary measurements}\label{appendix:characterization}

\subsection{Single-photon absorption line}\label{ss:onephotonline}
\begin{figure}[b]
    \centering \includegraphics[width=0.9\columnwidth]{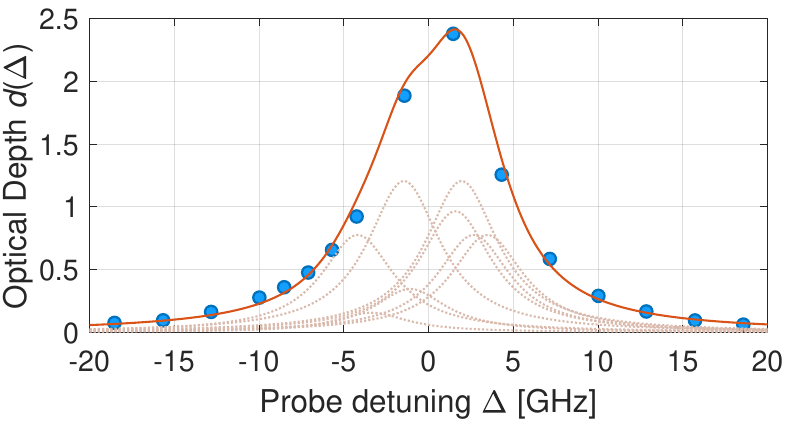}
    \caption{Single-photon absorption spectrum of the rubidium vapor cell with natural isotopic abundance.
    The measured optical depths (blue circles) are extracted from the logarithm of the transmitted power.
    The numerical fit (solid red line) is a sum of absorption profiles corresponding to the various hyperfine levels (dotted lines), all assumed to share a common homogeneous linewidth $\Gamma$. 
    From this fit, we extract $\Gamma=\unit[2.6]{GHz}$.
    Note that this measurement was performed at a reduced temperature relative to the main CPSR experiments.}
    \label{fig:absorption}
\end{figure}

Coherent polarization self-rotation is a two-photon effect that acts on the atomic spin in the vicinity of a single-photon optical transition.
We directly measure the optical absorption line and extract its homogeneous linewidth $\Gamma$, which is later used in the theoretical modeling.
This measurement uses a weak, linearly-polarized beam (absent the polarization modulation) whose frequency is scanned by varying the laser diode temperature.
Absorption data acquired at a low vapor temperature (to avoid absorption saturation) are plotted in Fig.~\ref{fig:absorption}.

We fit the data to a model that includes all hyperfine levels of naturally abundant rubidium within the D1 manifold~\cite{Siddons2008RbSpectroscopy,SteckRbData,Happer2010book}.
The transition frequencies and relative strengths are fixed to their known values, and we fit only two parameters: the overall optical depth and a common homogeneous linewidth $\Gamma$.
From this fit, we extract $\Gamma=\unit[2.6]{GHz}$,
consistent with pressure broadening due to 450 Torr of neon and 40 Torr of nitrogen buffer gases.
For optical fields far detuned from resonance, the ensemble response effectively reduces to that of a single transition with this homogeneous linewidth. 

\subsection{\label{ss:OP}Effect of optical pumping}

The performance of CPSR depends critically on the macroscopic axial spin polarization $p_\mathrm{a}>0$ maintained during operation.
This dependence is most evident in the linear scaling of the spin-light coupling rate $\Omega$ with $p_\mathrm{a}$. 
While the probe beam can coherently interact with the transverse spin component, it does not generate axial spin polarization.
To achieve and sustain the required polarization, CPSR relies on an auxiliary optical pumping beam.  

However, optical pumping also induces spin decoherence due to scattering, especially at high intensities.
This results in a trade-off: increasing the pump power improves spin polarization, but also shortens the coherence time.
Consequently, the contrast of CPSR spectral lines is expected to peak at an intermediate pumping rate.

To study the dependence on optical pumping, we apply a magnetic field of $B=\unit[1.2]{mG}$ and measure the CPSR spectrum across a range of pumping powers.
Due to the dependence of the slowing down factor and pump-induced vector lightshift on the pumping power, the Larmor frequency $\oa$ ranged between \unit[290]{Hz} and \unit[400]{Hz}.
We also increase the probe's single-photon detuning to $\Delta=115$~GHz to reduce probe absorption, at the expense of lower line contrast.
To quantify the optical pumping rate $\Rp/q$ for each power setting, we extract its contribution to the spin decoherence rate $\gamma(\Rp)$ from independent spin coherence-time measurements.
These measurements also allow us to determine the intrinsic (unpumped) spin decoherence rate by extrapolating to $\Rp=0$, yielding $\gamma(0)=21.5$ Hz.

\begin{figure}[!h]
    \centering \includegraphics[width=0.9\columnwidth]{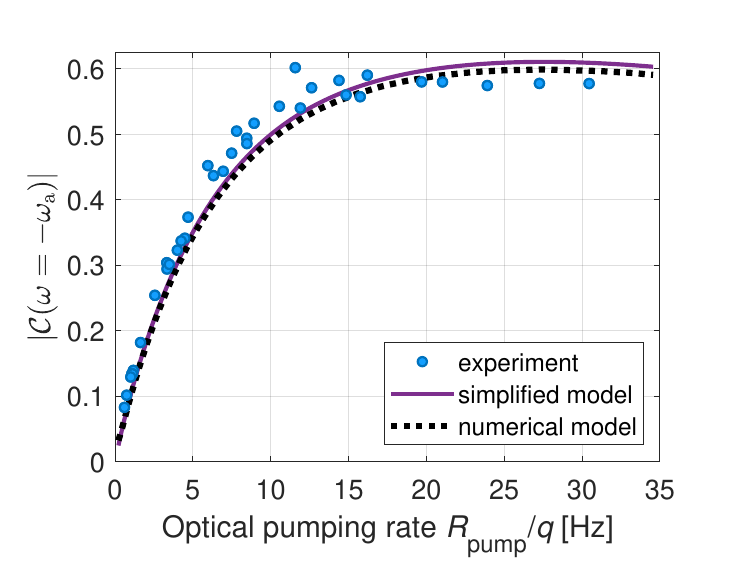}
    \caption{Signal amplification, given by $S_2^\text{out}e^{d(\Delta)}/S_2^\text{in}-1$ at the anti-resonance of CPSR, as a function of the optical pumping rate $\Rp/q$.
    The value of $\Rp/q$ is independently extracted from time-domain measurements of the spin decoherence rate.
    As optical pumping increases both spin polarization and decoherence, the amplification exhibits a maximum, observed when $\Rp/q=27.5$ Hz is comparable to the intrinsic spin decoherence rate $\gamma(\Rp=0)=21.5$ Hz.
    The solid purple line is the prediction of the simplified model, and the dashed black line is that of the calibrated numerical model.
    }
    \label{fig:OP}
\end{figure}

Figure \ref{fig:OP} shows the extracted line contrast as a function of the optical pumping rate $\Rp/q$.
As expected, the contrast reaches a maximum near $\Rp/q \approx \gamma(0)$, reflecting the trade-off between increased spin polarization and pump-induced decoherence.
Also shown are numerical results from our detailed theoretical model (Appendix\,\ref{sec:model}), which quantitatively capture this trade-off and show excellent agreement with the experimental data.

\section{\label{sec:model}Theoretical model}

This appendix presents a detailed theoretical model of coherent polarization self-rotation (CPSR). 
Table~\ref{tab:parameters} summarizes the variables and parameters used throughout.
The model, provided in subsection \ref{ss:detailedmodel}, extends beyond the simplified treatment in the main text: it includes the dynamics of the axial spin component, allows for large transverse spin components, and accounts for other non-idealities such as residual optical pumping induced by the probe 
and the spatial variation of both the spin and probe polarizations.
In subsection \ref{ss:simplemodel}, we derive the simplified model presented in Eqs.~(\ref{eq:in-out-model-1}) and (\ref{eq:in-out-model-2}) of the main text, and discuss the assumptions required for its validity.
The parameters used in the numerical solutions are provided in subsection \ref{ss:numerical}.

As CPSR implements a coherent interaction between the collective spin and optical fields, the same formalism applies when these fields are promoted to quantum operators.
This makes CPSR applicable to nonclassical states and quantum correlations.
In subsection \ref{subsec:operators}, we demonstrate this explicitly by formulating a quantum Heisenberg-Langevin description of the system. 

\subsection{Detailed model}\label{ss:detailedmodel}

The detailed model is based on Bloch equations, incorporating the relevant physical processes illustrated in Fig.~\ref{fig:apparatus}c~\cite{Happer2010book,Seltzer2008Deissertation}.
The excited-state population is adiabatically eliminated, assuming a predominant ground-state population.
Frequent spin-exchange collisions among alkali atoms drive this population towards a spin-temperature distribution, characterized by the spin polarization $p_\mathrm{a}$ and the slowing-down factor $q$~\cite{Walker1997SEOPReview,Happer2010book,AppeltHapper1998SEOPtheoryPRA}.
These same collisions are also responsible for the emergence of spin-exchange relaxation-free (SERF) behavior.

Under these conditions,  the coupled dynamics of the atomic spin and the optical probe field are given by
\begin{subequations} \label{eq:detailed-model}
\begin{align}
    \partial_t \mathbf{f}   =& -\oa \hat{\mathbf{z}}\times\mathbf{f} -\gamma(f_x \hat{\mathbf{x}}+f_y \hat{\mathbf{y}}) - \gamma_1 f_z \hat{\mathbf{z}} \label{eq:detailed-model-a} \\
    & + \frac{2}{q}\frac{\Delta}{\Gamma}\sigma(\Delta) s_3 \hat{\mathbf{x}}\times \mathbf{f} + \na( \frac{\Rp}{2}\hat{\mathbf{z}} + \sigma(\Delta) s_3\hat{\mathbf{x}} ),  \nonumber \\
    \partial_x \mathbf{s}  =& -\na \sigma(\Delta) \mathbf{s} - \frac{2}{q}\frac{\Delta}{\Gamma}\sigma(\Delta) f_x \left( \begin{array}{ccc} 0 & 1 & 0 \\ -1 & 0 & 0 \\ 0 & 0 & 0 \end{array} \right) \mathbf{s},  \label{eq:detailed-model-b}
\end{align}
\end{subequations}
where $\mathbf{f}$ and $\mathbf{s}$ -- the local spin densities and optical Stokes parameters --  as well as $\oa$ and $\Rp$, may vary spatially and temporally.
The spin relaxation terms $\gamma$ and $\gamma_1$ account for atomic diffusion and spin-destructive processes~\cite{Shaham2020Diffusion}.
All parameters are defined in Table~\ref{tab:parameters}.

\paragraph*{Isotope mixtures.}
For naturally abundant \textbf{rubidium}, we treat the mean-field spin as an effective composite spin comprising 72.2\% $^{85}\mathrm{Rb}$ and 27.8\% $^{87}\mathrm{Rb}$.
This treatment is justified by the rapid spin-exchange collisions between all rubidium atoms in the regime $R_\mathrm{se}\gg \oa,\omega$, which equilibrate the spin temperature (and thus the spin polarization $p_\mathrm{a}$) across both species.
As a result, the effective slowing-down factor is given by $q(p_\mathrm{a})=0.722q_{I=5/2}(p_\mathrm{a})+0.278q_{I=3/2}(p_\mathrm{a})$, ranging between $5.44=q(1)<q(p_\mathrm{a})<q(0)=10.8$, and the system can be described by a single coherent spin field $\mathbf{f}$.
For \textbf{potassium}, $q=q_{I=3/2}$ since the nuclear spin of the naturally abundant composition is almost entirely $I=3/2$.

\begin{table*}
\caption{\label{tab:parameters}Definitions of the model parameters.}
\resizebox{\textwidth}{!}{
\begin{tabular}{|c | c | c|}
\hline 
Notation & Description & Units \\ \hline \hline 
$\Na$ & Number of alkali atoms in the probe beam. & unitless \\
$\ell$ & Cell length. & $\unit{cm}$ \\ \hline
$A$ & Beam cross-sectional area, equal to the cell cross-section. The cell's volume is $A\ell$. & $\unit{cm^2}$ \\ \hline 
$T$ & Cell temperature. & Kelvin \\ \hline 
$\na$ & \parbox{0.75\textwidth}{Atomic number density $\na=\Na/(A\ell)$. For rubidium, $\na(T) [1/\text{cc}]= 10^{26.178 - 4040[\mathrm{K}]/T[\mathrm{K}]}/T[\mathrm{K}]$.  For potassium, $\na(T) [1/\text{cc}]= 10^{26.268 - 4453[\mathrm{K}]/T[\mathrm{K}]}/T[\mathrm{K}]$.} & $\unit{1/cm^3}$  \\ \hline 
$\Gamma$ & \parbox{0.75\textwidth}{Homogeneous broadening of the atomic optical linewidth, including buffer-gas induced pressure broadening. $2\Gamma$ is the D1 transition's FWHM.}& $\unit{GHz}$ \\ \hline 
$\Delta$ & Single-photon detuning: detuning of the probe from the alkali optical D1 transition. & $\unit{GHz}$ \\ \hline 
$\sigma(\Delta)$ & \parbox{0.75\textwidth}{Absorption cross-section of probe light $\sigma(\Delta)=( c r_e f_\text{osc} /\Gamma)/[1+(\Delta/\Gamma)^2]$, where $r_e$ the classical electron radius and $f_\text{osc}=0.34$ the oscillator strength for the alkali D1 transition.}& $\mathrm{cm^2}$ \\ \hline 
$d(\Delta)$ & Optical depth for the probe light in the cell, $d(\Delta)=\na\sigma(\Delta)\ell$. & unitless \\ \hline 
$p_\mathrm{a}$ & Alkali-spin polarization $-1\le p_\mathrm{a} \le 1$. & unitless \\ \hline 
$\oa$ & Larmor frequency, governed by the magnetic field and pump-induced vector lightshift. & Hz \\ \hline 
$q$ & \parbox{0.75\textwidth}{Slowing-down factor for the alkali spins (assuming a spin-temperature distribution),  $q=0.278q_{I=3/2}+0.722q_{I=5/2}$ with $q_{I=3/2}=\frac{6+2p_\mathrm{a}^2}{1+p_\mathrm{a}^2}$ and $q_{I=5/2}=\frac{38+52p_\mathrm{a}^2+6p_\mathrm{a}^4}{3+10p_\mathrm{a}^2+3p_\mathrm{a}^4}$ for naturally abundant rubidium, and $q=q_{I=3/2}$ for potassium~\cite{Katz2015SERFHybridization,Wei2020RbSEComag}.}& unitless \\ \hline 
$R_\mathrm{se}$ & \parbox{0.75\textwidth}{Spin-exchange collision rate. $R_\mathrm{se}=\na \sigma_\mathrm{se} v_T$, where $\sigma_\mathrm{se}=\unit[1.8\cdot10^{-14}]{cm^2}$ is the spin-exchange collision cross-section, and $v_T=\sqrt{8 k_B T/\pi \mu}$ is the mean thermal velocity for the colliding alkali atoms, with reduced mass $\mu=\unit[7.1\cdot10^{-23}]{gr}$ for rubidium and $\mu=\unit[3.25\cdot10^{-23}]{gr}$ for potassium.}& Hz \\ \hline 
$\gamma_\mathrm{se}$ & \parbox{0.75\textwidth}{Spin decoherence due to spin-exchange collisions in the SERF regime ($R_\mathrm{se}\gg \oa$), given by $\gamma_\mathrm{se}=[q^2-q(p_\mathrm{a}=1)^2]\oa^2/(2qR_\mathrm{se})$~\cite{Allred2002RomalisSERFmagnetometer}.} & Hz \\ \hline 
$\Rp$ & \parbox{0.75\textwidth}{Rate of optical pumping by the auxiliary pumping beam, given by $\Rp=\frac{\sigma(\Delta_\text{pump}) P_\text{pump}}{A(hc/\lambda)}$, where $\Delta_\text{pump}$ and $P_\text{pump}$ are the one-photon detuning and power of the pumping field. For hyperfine spin levels ($\hat{F}_z$-eigenstates) the effective rate is slowed down to $\Rp/q$.} & Hz\\ \hline 
$R_\mathrm{sd}$ & \parbox{0.75\textwidth}{Spin destruction rate, incorporating inter-atomic collisions, probe absorption, collisions with cell walls, and thermal atomic motion.} & Hz\\ \hline 
$\gamma_1$ & Spin depopulation rate, $\gamma_1=(R_\mathrm{sd}+\Rp)/q$. & Hz \\ \hline 
$\gamma$ & \parbox{0.75\textwidth}{Total spin decoherence rate, $\gamma = \gamma_1 + \gamma_\mathrm{se}$, including relaxations due to spin destruction, optical pumping, absorption of probe light, spin-exchange collisions (we assume no contribution from inhomogeneous Larmor precession).} & Hz \\ \hline 
$\hat{\mathbf{F}}_i$ & Hyperfine spin operators of the $i^\text{th}$ alkali atom in the ensemble. & unitless \\ \hline 
$\hat{\mathbf{F}}$ & Collective hyperfine spin operators of the alkali ensemble, $\hat{\mathbf{F}} = \sum_{i=1}^{\Na}\hat{\mathbf{F}}_i$. & unitless \\ \hline 
$\hat{\mathbf{f}}$ ; $\mathbf{f}=\langle\hat{\mathbf{f}}\rangle$ & \parbox{0.75\textwidth}{Local hyperfine spin-density operators and their expectation values, respectively. Formal definition given in Ref.~\cite{Shaham2020Diffusion}. For uniform spin fields, $\hat{\mathbf{f}} \sim \hat{\mathbf{F}}/(A\ell) \sim \na \hat{\mathbf{F}}_i$.} & $\mathrm{1/cm^3}$ \\ \hline 
$\Pc~;~\Ps$ & Optical power of the control and signal modes comprising the probe beam, $\Pc\gg\Ps$. & mW \\ \hline 
$S_{\{1,2,3\}}^\text{in/out}$ & \parbox{0.75\textwidth}{Incoming/outgoing Stokes parameters of the polarization-modulated probe beam. $S_1^\text{in}=\Pc/(2hc/\lambda)$ is approximately the probe photon flux. $S_\perp^\text{in} = \sqrt{|S_2^\text{in}|^2+|S_3^\text{in}|^2} = \sqrt{\Pc\Ps}/(hc/\lambda)$.} & Hz  \\ \hline 
$s_{\{1,2,3\}}^\text{in/out}$ & Local Stokes parameters. For a uniform probe, $s_\mu^\chi = S_\mu^\chi/A$.  & $\mathrm{Hz/cm^2}$ \\ \hline 
$\omega$ & Probe's polarization modulation frequency (control--signal modes two-photon detuning). & Hz  \\ \hline  
\end{tabular}
}
\end{table*}

\subsection{Simplified model and analytical solution}\label{ss:simplemodel}
The essence of CPSR is captured by a simplified model, derived by assuming uniformity of the pumping rate, Larmor frequency, and probe beam intensity over the cell volume, as well as negligible optical pumping by the probe, as detailed below. 
First, we assume that both the optical pumping rate $\Rp$ and the Larmor frequency $\oa$ are spatially homogeneous across the vapor cell.
Second, we assume that the probe beam uniformly covers the cell's cross-section and that its attenuation along the propagation axis is weak, \textit{i.e.} the off-resonant optical depth satisfies $d(\Delta)=\ell\na\sigma(\Delta)\ll1$.
Under these conditions, we directly solve Eq.~(\ref{eq:detailed-model-b}) for $s_3$, obtaining $s_3(x,t) = e^{-d(\Delta) x/\ell}s_3^\text{in}(t)$, compute its spatial average $\frac{1}{\ell}\int_0^\ell s_3(x,t)dx =\frac{1-e^{-d(\Delta)}}{d(\Delta)}s_3^\text{in}$, and substitute this into Eq.~(\ref{eq:detailed-model-a}).
As a result, Eqs.~(\ref{eq:detailed-model-a}) reduce to purely temporal equations, and the spin field $\mathbf{f}$ becomes spatially uniform.
Solving the resulting system for $s_{1,2}$, we obtain the following dynamical input-output relations:
\begin{subequations} \label{eq:detailed-in-out-model}
\begin{align}
    \partial_t \mathbf{f}   =& -\oa \hat{\mathbf{z}}\times\mathbf{f} -\gamma(f_x \hat{\mathbf{x}}+f_y \hat{\mathbf{y}}) - \gamma_1 f_z \hat{\mathbf{z}} \nonumber \\
    & + \frac{2}{q}\frac{1-e^{-d(\Delta)}}{d(\Delta)}\frac{\Delta}{\Gamma}\sigma(\Delta) s_3^\text{in} \hat{\mathbf{x}}\times \mathbf{f}\nonumber \\
    &  + \na( \frac{\Rp}{2}\hat{\mathbf{z}} + \frac{1-e^{-d(\Delta)}}{d(\Delta)}\sigma(\Delta) s_3^\text{in}\hat{\mathbf{x}} ), \label{eq:D2a}\\
    s_1^\text{out} =& e^{-d(\Delta)}[ \cos{(2\theta)}s_1^\text{in} - \sin{(2\theta)}s_2^\text{in} ], \\
    s_2^\text{out} =& e^{-d(\Delta)}[ \cos{(2\theta)}s_2^\text{in} + \sin{(2\theta)}s_1^\text{in} ], \\
    s_3^\text{out} =& e^{-d(\Delta)}s_3^\text{in},
\end{align}
\end{subequations}
where $\theta=\tfrac{\Delta}{\Gamma}\sigma(\Delta) \ell \tfrac{f_x}{q}$ is the Faraday rotation angle.

At this point, we assume the CPSR regime in which optical pumping by the probe beam is negligible, and omit the term proportional to $\sigma(\Delta) s_3^\text{in}\hat{\mathbf{x}}$ in Eq.~(\ref{eq:D2a}).
This assumption holds when the auxiliary pump dominates the pumping process, $\Rp\gg \sigma(\Delta) s_3$, a condition that is facilitated when the probe is both strongly detuned $\Delta\gg \Gamma$ and predominantly linearly polarized ($\Pc\gg \Ps$).
The spin polarization is then maintained primarily by the auxiliary pump beam, in contrast to conventional PSR, where the spin polarization is generated directly by the elliptically polarized probe near resonance.

It follows that $f_{x,y} \ll f_z=  \na q p_\mathrm{a}/2$, where the axial spin polarization is given by $p_\mathrm{a} = \Rp/q\gamma_1 = \Rp/(\Rp+R_\mathrm{sd})$, and therefore $\theta \ll 1$.
The equations then simplify to:
\begin{subequations} \label{eq:HP-model}
\begin{align}
    \partial_t f_x  & = -\gamma f_x +\oa f_y, \label{eq:HP-model-a} \\
    \partial_t f_y  & = -\oa f_x -\gamma f_y -     (k_{{\mathrm{L}\rightarrow\mathrm{a}}}/\ell)s_3^\text{in} , \label{eq:HP-model-b} \\
    f_z  & = q \na p_\mathrm{a}/2, \label{eq:HP-model-c} \\
    s_1^\text{out} &= e^{-d(\Delta)}s_1^\text{in}, \label{eq:HP-model-d} \\
    s_2^\text{out} &= e^{-d(\Delta)}(s_2^\text{in} +     k_{{\mathrm{a}\rightarrow\mathrm{L}}} f_x \ell
    ), \label{eq:HP-model-e} \\
    s_3^\text{out} &= e^{-d(\Delta)}s_3^\text{in}. \label{eq:HP-model-f}
\end{align}
\end{subequations}
where the coupling constants, as defined in the main text, are
\begin{equation}
k_{{\mathrm{L}\rightarrow\mathrm{a}}}= \frac{1-e^{-d(\Delta)}}{d(\Delta)} \frac{2}{q} \frac{\Delta}{\Gamma} \sigma(\Delta)  f_z \ell   \mathrm{~,~} k_{{\mathrm{a}\rightarrow\mathrm{L}}}= \frac{2}{q} \frac{\Delta}{\Gamma} \sigma(\Delta) s_1^\text{in}.
\end{equation}
We recover Eqs.~(\ref{eq:in-out-model-1}) and (\ref{eq:in-out-model-2}) in the main text by integrating Eqs.~(\ref{eq:HP-model-a})-(\ref{eq:HP-model-c}) over the cell volume and integrating Eqs.~(\ref{eq:HP-model-d})-(\ref{eq:HP-model-f}) over the cell cross-section. For uniform fields, this simply amounts to substituting $\mathbf{f}=\mathbf{F}/(A\ell)$ and   $\mathbf{s}=\mathbf{S}/A$.

To derive the CPSR spectrum, we now introduce the polarization modulation of the probe, with a frequency offset $\omega$ between $\mathbf{s}_{2,3}$ (signal field) and $\mathbf{s}_1$ (control field): $\mathbf{s}^\text{in/out}=(s_1^\text{in/out},s_\perp^\text{in/out} e^{i\omega t},i s_\perp^\text{in/out} e^{i\omega t})^T$.
We transform Eqs.~(\ref{eq:HP-model-a}) and (\ref{eq:HP-model-b}) to Fourier space,
\begin{align}\label{eq:fourier-model}
    i\omega f_x(\omega)   =& -\gamma f_x(\omega) +\oa f_y(\omega), \\
    i\omega f_y(\omega)   =& -\oa f_x(\omega) -\gamma f_y(\omega) 
     - (k_{{\mathrm{L}\rightarrow\mathrm{a}}}/\ell)s_3^\text{in}(\omega), \nonumber
\end{align}
using $s_2^\text{in/out}(\omega) = -i s_3^\text{in/out}(\omega)= s_\perp^\text{in/out}(\omega)$, to find
\begin{align}
    f_x & = \frac{k_{{\mathrm{L}\rightarrow\mathrm{a}}}}{\ell}
    \frac{-i \oa}{(\gamma+i\omega)^2 + \oa^2} s_2^\text{in}, \nonumber \\
    f_y  & = \frac{k_{{\mathrm{L}\rightarrow\mathrm{a}}}}{\ell}
    \frac{-i(\gamma+i\omega)}{(\gamma+i\omega)^2 + \oa^2} s_2^\text{in}, 
\end{align}
and, with $\Omega\equiv k_{{\mathrm{L}\rightarrow\mathrm{a}}}k_{{\mathrm{a}\rightarrow\mathrm{L}}}$, obtain
\begin{equation}\label{eq:analytical-solution}
    \frac{s_2^\text{out}}{e^{-d(\Delta)}s_2^\text{in}} = 1 - \frac{i\Omega\oa}{\oa^2-(\omega-i\gamma)^2}. 
\end{equation}
Equation (\ref{eq:analytical-solution}) yields Eq.~(\ref{eq:spectrum}) in the main text, with  $S_{1,2}^\text{in/out} = A s_{1,2}^\text{in/out}$.

\subsection{Numerical solution}\label{ss:numerical}
The theoretical curves shown in Figs. \ref{fig:spectrum}, \ref{fig:SERF}, \ref{fig:10Hz}, and \ref{fig:OP}. 
For the simplified model [Eq.~(\ref{eq:spectrum})], we derive $\Omega,\oa,\gamma$ from the model's parameters.
For the numerical model, solutions are obtained by numerically solving Eqs.~(\ref{eq:detailed-in-out-model}).
The simulations assume a spatially uniform optical pumping rate $\Rp$ and Larmor frequency $\oa$, as well as a probe beam that uniformly illuminates the entire cross-sectional area of the cell.
For rubidium, natural isotopic abundance is used.
The parameters used in these calculations are listed in Table~\ref{tab:values}.

For the calculations of the rubidium experiments in Figs. \ref{fig:spectrum}, \ref{fig:SERF}, and \ref{fig:OP}, the model parameters yield a bare spin decoherence rate of $\gamma(\Rp=\oa=0)=13.2$~Hz.
The discrepancy with the experimentally measured value (21.5~Hz) may arise from spatial inhomogeneities in the pumping rate and lightshifts, caused by partial depletion of the pumping beam.

\begin{table}[!tb]
\resizebox{\columnwidth}{!}{
\begin{tabular}{|c|c|c|c|}
\hline
Parameter & \multicolumn{1}{c|}{Calculations for} & \multicolumn{1}{c|}{Calculation for} & \multicolumn{1}{c|}{Calculation for} \\
 & \multicolumn{1}{c|}{Figs.~\ref{fig:spectrum}, \ref{fig:SERF} \hspace{0.5cm}} & \multicolumn{1}{c|}{Fig.~\ref{fig:OP} \hspace{0.5cm}} & \multicolumn{1}{c|}{Fig.~\ref{fig:10Hz} \hspace{0.5cm}} \\ \hline
\hline 
$\ell$ & \multicolumn{2}{c|}{1~cm} & \parbox{2.4cm}{1~cm (effective for a 14~mm diameter sphere)} \\ \hline 
$A$ & \multicolumn{2}{c|}{1~cm$^{2}$} & 1.5~cm$^2$ \\ \hline 
$\Gamma$ & \multicolumn{2}{c|}{2.6~GHz} & 12~GHz \\ \hline 
$\Delta_{\text{pump}}$ & \multicolumn{2}{c|}{45~GHz} & 100~GHz \\ \hline 
$\Pc$ & \multicolumn{2}{c|}{15.2~mW} & 25~mW \\ \hline 
$\Ps$ & \multicolumn{2}{c|}{20~$\mu$W} & 35~$\mu$W \\ \hline 
$P_\text{pump}$ & 60~mW & 0--100~mW & \parbox{2.4cm}{1.6~mW (effective over the cell)} \\ \hline \hline 
$\oa$ & 268~Hz (Fig.~\ref{fig:spectrum}) & 290--400~Hz & 29~Hz \\ \hline 
$T$ & \multicolumn{2}{c|}{154~$^\circ$C} & 185~$^\circ$C \\ \hline 
$\Delta$ & 89~GHz & 116~GHz & 215~GHz \\ \hline 
$R_\mathrm{sd}$ & 98~Hz & 94~Hz & 27~Hz \\ \hline 
\end{tabular}
}
\caption{\label{tab:values}
Parameter values used in the calculations (with $\mathrm{Hz}\equiv2\pi\cdot\mathrm{s}^{-1}$). The parameters $\oa$, $T$, $\Delta$, and $R_\mathrm{sd}$ were fine-tuned within the experimental uncertainty (note that $R_\mathrm{sd}$ includes probe absorption, which is higher than in the pumping characterization experiments).}
\end{table}

\subsection{\label{subsec:operators}Simplified Heisenberg-Langevin model for quantum operators}

To extend the CPSR framework to the quantum regime, we consider the action of the dispersive Faraday Hamiltonian on the quantum operators of the collective atomic spin and the optical probe field.
This Hamiltonian is valid when the spins are transversely polarized and the probe is far-detuned from the optical transition.
It describes both the vector light-shift and Faraday rotation that underlie CPSR.
We derive the corresponding Heisenberg-Langevin equations, which capture both the coherent dynamics and quantum noise contributions.
The resulting equations closely resemble their classical counterparts, \textit{e.g.,} Eqs.~(\ref{eq:in-out-model-1}) and (\ref{eq:in-out-model-2}) from the main text, but are expressed directly in terms of operators rather than expectation values.
This formalism shows that CPSR can support nonclassical states and correlations, enabling applications in quantum optics and quantum information processing.

We describe the optical field using the quantum Stokes operators, defined as~\cite{Polzik2010ReviewRMP}
\begin{subequations} \label{eq:nonclassical-stokes}
\begin{align}
    \hat{s}_1(x) & = \tfrac{1}{2}[\hat{a}^\dagger_z(x)\hat{a}_z(x) - \hat{a}^\dagger_y(x)\hat{a}_y(x)], \\
    \hat{s}_2(x) & = \tfrac{1}{2}[\hat{a}^\dagger_z(x)\hat{a}_y(x) + \hat{a}^\dagger_y(x)\hat{a}_z(x)], \\
    \hat{s}_3(x) & = \tfrac{1}{2i}[\hat{a}^\dagger_z(x)\hat{a}_y(x) - \hat{a}^\dagger_y(x)\hat{a}_z(x)],
\end{align}
\end{subequations}
where $\hat{a}_z(x)$ and $\hat{a}_y(x)$ are the annihilation operators for the $z$-polarized control and $y$-polarized signal modes, respectively, at spatial position $x$.
These operators satisfy the bosonic commutation relations $[\hat{a}_\chi(x),\hat{a}^\dagger_{\chi'}(x')]=c\delta_{\chi,\chi'}\delta(x-x')$, and $\langle\hat{a}^\dagger_{\chi}(x)\hat{a}_{\chi}(x)\rangle$ is the local photon flux of the mode $\chi\in \{y,z\}$.
The classical Stokes parameters correspond to the expectation values of the quantum Stokes operators.

Under the Holstein-Primakoff approximation for the optical field~\cite{Polzik2010ReviewRMP}, the control mode is treated as a strong, classical field, while the signal mode remains a weak quantum field.
This allows the quantum operators associated with the control to be replaced by classical variables.
In particular, the operator $\hat{s}_1$ is replaced by its expectation value $\hat{s}_1\mapsto \langle \hat{s}_1 \rangle=s_1$.
Similarly, we apply the Holstein-Primakoff approximation to the spin density operators $\hat{f}_x$, $\hat{f}_y$, and $\hat{f}_z$.
Due to the optical pumping, $\hat{f}_z$ can be treated as a macroscopic, classical number $\hat{f}_z\mapsto \langle \hat{f}_z \rangle=f_z$.
Under these approximation, the commutation relations for the light and spin operators become $[\hat{s}_2(x),\hat{s}_3(x')]=ics_1(x)\delta(x-x')$ and $[\hat{f}_x(\mathbf{r}),\hat{f}_y(\mathbf{r}')]=if_z(\mathbf{r})\delta(\mathbf{r}-\mathbf{r}')$.

The Faraday interaction Hamiltonian for far-detuned light propagating through a spin ensemble in a spin-temperature distribution is given by
\begin{equation}
    \mathcal{H} = \hbar\int_\text{volume}d^3r \frac{2}{q} \frac{\Delta}{\Gamma}\sigma(\Delta) \hat{s}_3(\mathbf{r}) \hat{f}_x(\mathbf{r}).
\end{equation}
Including the magnetic (Larmor precession) Hamiltonian, spin relaxation, light absorption, and associated fluctuations, and using $\partial_x \hat{s}_j = -i[\hat{s}_j,\mathcal{H}]/c$, we derive the Heisenberg-Langevin equations for the local spin and Stokes operators:
\begin{align}\label{eq:nonclassical-FPOR}
    \partial_t \hat{f}_x & = -\gamma \hat{f}_x +\oa \hat{f}_y + \hat{f}_\text{spin-noise}^{(x)}, \nonumber \\
    \partial_t \hat{f}_y & = -\oa \hat{f}_x -\gamma \hat{f}_y - \frac{2}{q} \frac{\Delta}{\Gamma} \sigma(\Delta) f_z \hat{s}_3 + \hat{f}_\text{spin-noise}^{(y)},\nonumber  \\
    \partial_x \hat{s}_2 & = -\na \sigma(\Delta) \hat{s}_2 + \frac{2}{q} \frac{\Delta}{\Gamma} \sigma(\Delta) s_1 \hat{f}_x + \hat{f}_\text{photon-noise}^{(2)}, \nonumber \\
    \partial_x \hat{s}_3 & = -\na \sigma(\Delta) \hat{s}_3 + \hat{f}_\text{photon-noise}^{(3)},
\end{align}
The quantum noise terms arise from the fluctuation-dissipation relation and ensure preservation of the canonical commutation relations.

To elucidate the nonclassical dynamics, we recast Eqs.~(\ref{eq:nonclassical-FPOR}) in terms of the uniform spin and optical mode operators.
We define the photon flux in the control field $\Phi_c$, and express the Stokes operators as $s_1=\Phi_c/(2A)$, $\hat{s}_2=\sqrt{\Phi_c}\mathrm{Re}\,\hat{a}_y/A$, and  $\hat{s}_3=\sqrt{\Phi_c}\mathrm{Im}\,\hat{a}_y/A$.
The transverse spin components are normalized as $\hat{f}_{\mu}=\hat{F}_{\mu}/(A\ell)$ for $\mu\in\{x,y\}$, and we set $f_z=\na q p_\mathrm{a}/2$ as before.
Solving for the signal mode output, we obtain:
\begin{align} \label{eq:nonclassical-FPOR-photons}
    \partial_t \hat{F}_x & = -\gamma \hat{F}_x +\oa \hat{F}_y + \hat{\mathcal{F}}_\text{spin-noise}^{(x)}, \nonumber \\
    \partial_t \hat{F}_y & = -\oa \hat{F}_x -\gamma \hat{F}_yk_{{\mathrm{L}\rightarrow\mathrm{a}}}\sqrt{\Phi_c}\mathrm{Im}\,\hat{a}_y + \hat{\mathcal{F}}_\text{spin-noise}^{(y)}, \nonumber \\
    \hat{a}_y^\text{out} & = e^{-d(\Delta)} \left( \hat{a}_y^\text{in} + \frac{k_{{\mathrm{a}\rightarrow\mathrm{L}}}}{\sqrt{\Phi_c}}
    \hat{F}_x \right),
\end{align}
in direct analogy with the classical model in Eqs.~(\ref{eq:HP-model}).

\bibliography{CPSR}

\end{document}